\documentclass[twocolumn]{article}
\usepackage{parskip}
\usepackage{amssymb}
\usepackage{amsmath}
\usepackage{fullpage}
\usepackage{authblk}
\usepackage{blindtext}
\usepackage{siunitx}
\usepackage{float}
\usepackage{caption}
\usepackage[superscript,nomove]{cite}
\usepackage{hyperref}
\usepackage[normalem]{ulem}
\usepackage{graphicx}
\usepackage{cancel}
\usepackage{pifont}
\usepackage{diagbox}
\usepackage{tikz}

\makeatletter \renewcommand{\@citess}[1]{\textsuperscript{[#1]}} \makeatother

\makeatletter
\def\@firstoftwo@second#1#2{%
  \def\temp##1.##2\@nil{##2}%
   \temp#1\@nil}
\newcommand\sref[1]{%
   (A.\expandafter\@setref\csname r@#1\endcsname\@firstoftwo@second{#1})%
}

\makeatother

\title{A general statistical framework for vacancy and self-interstitial properties in concentrated multicomponent solids}
\author[1]{\small Jacob Jeffries \thanks{jwjeffr@g.clemson.edu}}
\author[2]{Hyunsoo Lee}
\author[3]{Anter El-Azab}
\author[1,2]{Enrique Martinez \thanks{enrique@clemson.edu}}
\affil[1]{Department of Materials Science and Engineering, Clemson University, Clemson, SC 29634, USA}
\affil[2]{School of Mechanical and Automotive Engineering, Clemson University, Clemson, SC 29634, USA}
\affil[3]{School of Materials Engineering, Purdue University, West Lafayette, IN 47907, USA}
\date{\small \today}

\renewenvironment{abstract}
 {\quotation\small\noindent\rule{\linewidth}{.5pt}\par\smallskip
  {\centering\bfseries\abstractname\par}\medskip}
 {\par\noindent\rule{\linewidth}{.5pt}\endquotation}

\begin{document}

\twocolumn[
  \begin{@twocolumnfalse}
  \maketitle
    \begin{abstract}
        A rigorous understanding of the thermodynamic properties of point defects, namely vacancies and self-interstitials, is crucial for the discovery and screening of structural materials in clean energy applications. In this work, we extend a previously-developed statistical framework for predicting the thermodynamics of single-site impurities to further predict the thermodynamics of self-interstitial dumbbells in an arbitrarily complex alloy. We then apply this extended framework to compute effective formation energies in fully disordered Fe-Cr and Cu-Ni alloys. Notably, we predict that some self-interstitial dumbbell types that are high-energy in pure Fe become stabilized by Cr. We additionally describe a symmetry-breaking effect, wherein high solute concentrations distort the defect free energy surface, yielding misaligned self-interstitials.
    \end{abstract}
  \vspace{0.5cm}
  \end{@twocolumnfalse}
]

\section{Introduction}

In nuclear environments, continuous bombardment by energetic neutrons causes material damage.\cite{bocci2020arc} 
Point defects (vacancies and self-interstitials) and their clusters are generated as neutrons displace atoms from their lattice sites.\cite{nordlund2018primary, nespolo2017international} The evolution of these defects is at the heart of the microstructure changes in the material, which alter the properties of the system. Consequently, the limits of material performance in these environments are governed by the physical and thermodynamic properties of Frenkel pairs and their constituent point defects (PD), rather than by ideal crystalline properties alone. \cite{lucas1994effects, lucas2007structure, chopra2011review}

In metallic systems, self-interstitial atoms (SIAs) generally diffuse much faster than vacancies and interact more strongly with microstructure features such as dislocations and grain boundaries. Higher mobility leads to a larger propensity for the formation of clusters that will generate dislocation loops. Such loops hinder plastic deformation leading to hardening and embrittlement in a wide range of temperatures. \cite{was2007fundamentals,zinkle2005advanced} Furthermore, SIA fluxes couple with fluxes of alloying elements to modify the chemical distribution in the material, leading to radiation-induced segregation (RIS) and precipitation (RIP), and chemical inhomogeneities that modify the material response and can lead to processes such as irradiation-assisted stress corrosion cracking.\cite{badillo2015phase, piochaud2016atomic,dubey2013irradiation, bellon2020phase} The basic mechanisms for SIA formation and migration depend on the details of the lattice and electronic structures. 

In pure $\alpha$-Fe with a body-centered cubic (BCC) crystalline structure, and induced by magnetic effects, a $\langle 110 \rangle$ dumbbell configuration is energetically favorable \cite{Fu2004} compared to the more common $\langle 111 \rangle$ crowdion in other BCC elements such as W \cite{nguyen2006self}. This difference in the ground states has profound implications in the microstructure evolution since in $\alpha$-Fe, SIAs diffuse in 3D through a Johnson mechanism \cite{Willaime2005}, while in W, SIA migrate mainly in 1D,\cite{Fitzgerald2008} modifying the reaction rates between SIAs and other defects. In face-centered cubic (FCC) metals, the most common ground state SIA configuration is the $\langle 100 \rangle$, although some exceptions have been observed with octahedral and $\langle 110 \rangle$ configurations also possible.\cite{PhysRevMaterials.5.013601} Therefore, understanding defect ground state configurations is paramount to predict material changes affecting its properties. Alloying modifies the ground state and the propensity for rotation to different states, which again modifies the diffusion characteristics and the evolution of the microstructure.\cite{watanabe2019tensile, zheng2023preparation, yin2021ductile, korpe2025effect, el2019outstanding}

Irreversible thermodynamic approaches are needed to predict the chemical redistribution of materials under irradiation.\cite{nastar20121} These models describe particle and defect fluxes as a function of thermodynamic driving forces and transport coefficients:
\begin{equation}
    \mathbf{J}_\alpha = -\sum_{\alpha'} L_{\alpha\alpha'} \nabla\mu_{\alpha'}
\end{equation}
where $\mathbf{J}_\alpha$ is the flux of species $\alpha$, $\mu_{\alpha'}$ is the chemical potential of species $\alpha'$, and $L_{\alpha\alpha'}$ is the \textit{equilibrium} $\alpha$-$\alpha'$ Onsager coefficient.\cite{PhysRev.38.2265} $L_{\alpha\alpha'}$ can be computed by inserting a single PD in a periodic box and analyzing the atomic trajectories using an atomistic method such as molecular dynamics \cite{doi:10.1080/08927022.2020.1810685} or atomistic kinetic Monte Carlo,\cite{PhysRevB.88.134207,B101982L,piochaud2016atomic} evaluating the atomic displacements through the generalized Einstein relation\cite{ARAllnatt_1982}:
\begin{equation}
    L_{\alpha\alpha'}^{(1)} = \frac{1}{Vk_BT}\lim_{t\to\infty}\frac{\left\langle \mathbf{R}^{(1)}_\alpha(t)\cdot \mathbf{R}^{(1)}_{\alpha'}(t)\right\rangle}{2dt}
\end{equation}
In this formulation, $V$ represents the system volume, $k_B$ is the Boltzmann constant, and $T$ is the temperature. The term $\mathbf{R}_\alpha^{(1)}(t)$  denotes the total displacement of atomic species $\alpha$ at time $t$ as mediated by a single PD, where $\left\langle\cdot\right\rangle$ signifies the ensemble average and $d$ the problem dimensionality. The resulting $\alpha$-$\alpha'$ transport coefficient, $L_{\alpha\alpha'}^{(1)}$, is an extensive quantity proportional to system size.

In contrast, the physical transport coefficient $L_{\alpha\alpha'}= nL_{\alpha\alpha'}^{(1)}$ is intensive, where $n$ represents the equilibrium defect population within a similar volume. Utilizing atomic-scale models to compute $L_{\alpha\alpha'}^{(1)}$ presents significant challenges: these simulations are computationally demanding and typically necessitate defect concentrations several orders of magnitude above thermodynamic equilibrium. Furthermore, extracting $L_{\alpha\alpha'}$ coefficients from experimental diffusion data—particularly those governed by SIA migration—remains exceptionally difficult. Consequently, accurate estimation of transport coefficients requires precise determination of the equilibrium PD concentration.

In this work, we extend a previously developed statistical framework for computing properties of single-site impurities like vacancies \cite{jeffries2025prediction} to predict properties of SIAs accounting for their configurational complexities. To efficiently capture such complexity, we identify equivalent defect microstates, i.e. configurations that share the same symmetry, allowing the model to describe the ensemble-averaged populations of point defects across the alloy. This statistical framework provides composition and temperature-dependent effective formation energies, which serve as input to irreversible thermodynamics models to predict microstructure evolution, including radiation-induced segregation and defect-driven compositional rearrangements.

From this framework, we compute composition and temperature-dependent effective formation energies for two alloys of interest: disordered $\text{Fe}_{1-x}\text{Cr}_x$ (BCC) and $\text{Cu}_{1-x}\text{Ni}_x$ (FCC), each with potentially different SIA ground state configurations. We additionally identify a symmetry-breaking effect of the point defect energy landscape, in which certain self-interstitial configurations become misaligned after relaxation, highlighting the significant effect of local chemical environments on both quantitative and qualitative properties of point defects.
\section{Model}

Within the framework previously developed by Jeffries et al. \cite{jeffries2025prediction}, we treat defects as low-probability microstates within the grand canonical ensemble, and compute the concentration of each defect type by computing an ensemble average over the solid solution's lattice sites, indexed by $\sigma$. Additionally, we only consider point defects, and save higher-dimensional defects for a future work.

In a homogeneous solid, the enumeration of possible defect microstates is trivial, namely the occupation by an atom, a vacancy, and any one of the self-interstitials of interest. For example, in pure Fe, the possible occupying types at any given lattice site are an iron atom, a vacancy, a $\langle 110 \rangle$ self-interstitial, and a $\langle 111 \rangle$ self-interstitial. In the case of a homogeneous medium, there is significant degeneracy in the dumbbell configurations, e.g. a $[110]$ self-interstitial is equivalent to a $[101]$ self-interstitial and a $[ 111]$ self-interstitial is equivalent to a $[11\overline{1}]$ self-interstitial.

However, in a disordered multicomponent solid, solid solution noise breaks the symmetry of the configuration space described above, lifting the degeneracy of the defect microstates, combinatorically exploding the number of possible unique microstates. Namely, for each lattice site $\sigma$, we have the following microstates:

\begin{itemize}
    \item Occupation by all elements of type $\alpha$
    \item Occupation by a vacancy
    \item Occupation by all $\alpha$-$\alpha'$ self-interstitials sitting along direction $\mathbf{n}$, where $\mathbf{n}$ is the vector pointing from the $\alpha$ atom to the $\alpha'$ atom, unique up to spatial symmetries
\end{itemize}

We respectively denote the energy of these microstates as $E_\sigma^{(\alpha)}$, $E_\sigma^{(v)}$, and $E_\sigma^{(\mathbf{n}, \alpha,\alpha')}$, which can be computed from a large class of atomistic methods. Then, in the grand canonical ensemble \cite{sadigh2012scalable}, the microstates of the first kind have probability:

\begin{equation}
    p_\sigma^{(\alpha)}(\beta) \sim \exp\left(-\beta \Tilde{E}_\sigma^{(\alpha)}\right)
\end{equation}

while the vacant microstate has probability:

\begin{equation}
    p_\sigma^{(v)}(\beta) \sim \exp\left(-\beta \Tilde{E}_\sigma^{(v)}\right)
\end{equation}

and the self-interstitial microstates have probability:

\begin{equation}
    p_\sigma^{(\mathbf{n}, \alpha,\alpha')}(\beta)\sim \exp\left(-\beta\Tilde{E}_\sigma^{(\mathbf{n}, \alpha,\alpha')}\right)
\end{equation}

where the $\Tilde{E}$'s denote the Legendre-transformed potential appropriate for the grand canonical ensemble, namely:

\begin{equation}
    \begin{aligned}
        \Tilde{E}_\sigma^{(\alpha)} &= E_\sigma^{(\alpha)} - \mu_\alpha \\
        \Tilde{E}_\sigma^{(v)} &= E_\sigma^{(v)} \\
        \Tilde{E}_\sigma^{(\mathbf{n}, \alpha,\alpha')} &= E_\sigma^{(\mathbf{n}, \alpha,\alpha')} - \mu_{\alpha} - \mu_{\alpha'}
    \end{aligned}
\end{equation}

where $\mu_\alpha$ is the chemical potential of alloying type $\alpha$, and $\beta = 1/(k_BT)$ is the inverse temperature, where $k_B = \SI{8.617e-5}{eV/K}$ is the Boltzmann constant and $T$ the temperature. Additionally, the probabilities $p$ are normalized to sum to unity:

\begin{equation}
    \sum_\alpha p_\sigma^{(\alpha)}(\beta) + p_\sigma^{(v)}(\beta) + \sum_{(\mathbf{n}, \alpha, \alpha')} p_\sigma^{(\mathbf{n}, \alpha, \alpha')}(\beta) = 1
\end{equation}

From these probabilities, we compute the vacancy concentration:

\begin{equation}\label{eq:vacancy-concentration}
    x_v(\beta) = \left\langle p_\sigma^{(v)}(\beta)\right\rangle
\end{equation}

as well as the concentration of a given $\alpha$-$\alpha'$ interstitial dumbbell along the $\mathbf{n}$ direction:

\begin{equation}
    x_{\mathbf{n}, \alpha, \alpha'}(\beta) = \left\langle p_\sigma^{(\mathbf{n}, \alpha, \alpha')}(\beta)\right\rangle
\end{equation}

\begin{figure}[H]
    \centering
    \includegraphics[width=\linewidth]{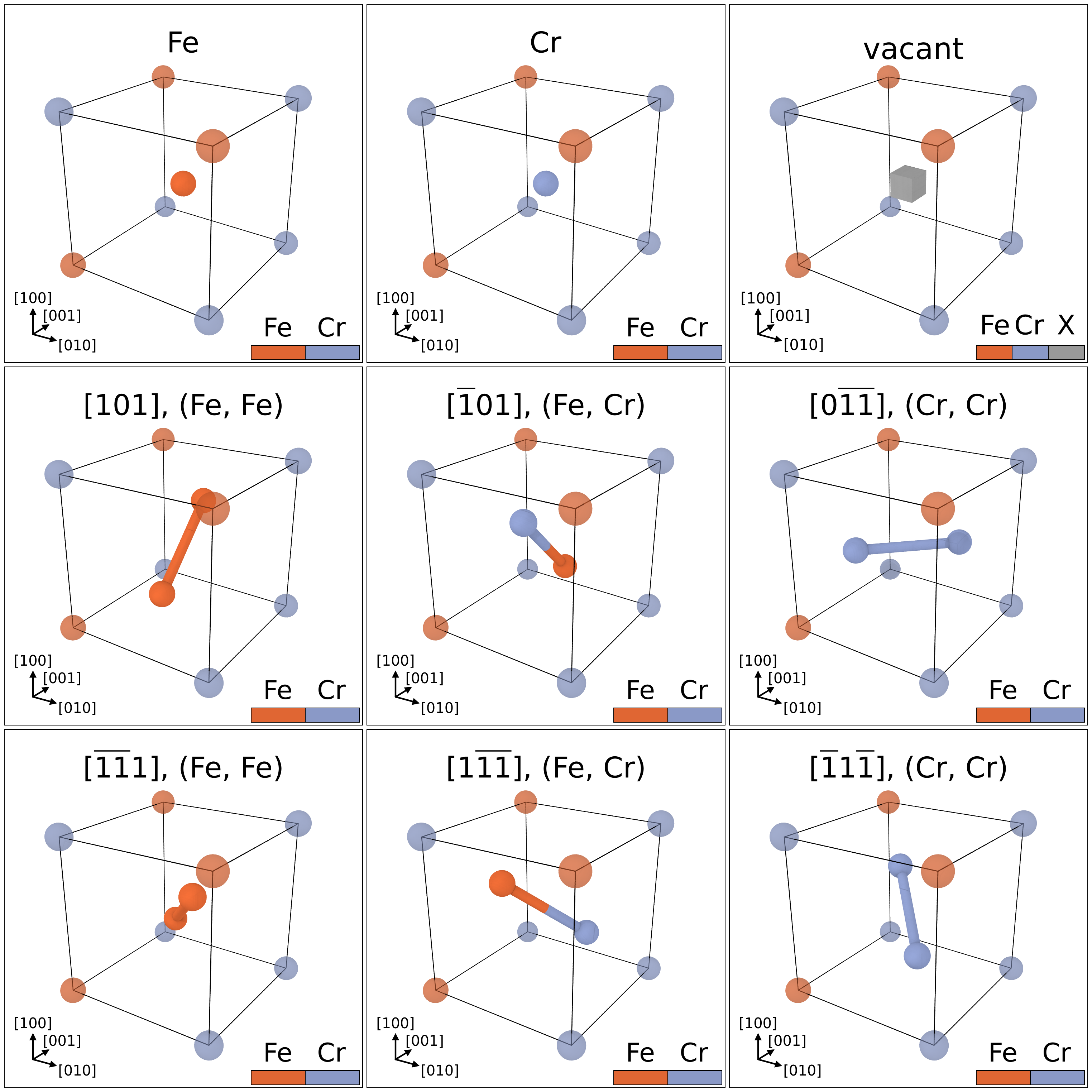}
    \caption{Example possible microstates in $\text{Fe}_{1-x}\text{Cr}_x$ visualized with Open Visualization Tool (OVITO) \cite{stukowski2009visualization}.}
    \label{fig:dumbbell-vis}
\end{figure}

This defines a set of unique microstates for a given local chemical environment. For the $\text{Fe}_{1-x}\text{Cr}_x$ binary system, for example, the possible directions $\mathbf{n}$ are in the $\langle 110\rangle$ and $\langle 111\rangle$ families, and the possible chemical types are $\{\alpha, \alpha'\} = \{\text{Fe}, \text{Cr}\}$. A subset of these microstates in an arbitrary local chemical environment is visualized in Fig.~\ref{fig:dumbbell-vis}. Note that, although only $9$ microstates are shown, there are $43$ total possible microstates due to the configurational complexity of the SIA microstates. A visualization of the full set of possible microstates is available in the Supplementary Materials.

However, these large sets of microstates can be reduced into sets of equivalent microstates. Namely, we are largely interested about the concentration of SIA within equivalence classes defined by symmetry, i.e. a $\left [ 110\right ]$ Fe-Cr dumbbell and a $\left [ \overline{1}10\right ]$ Fe-Cr dumbbell are often left undistinguished, both labeled as $\langle 110\rangle$ Fe-Cr dumbbells.

More generally, let $N$ be the set of lattice vectors that we wish to compute SIA concentrations of. For example, in a BCC solid, we choose $N = \{[110], [111]\}$ from \textit{a priori} knowledge about which SIA types tend to be metastable. Without much loss of generality, we will assume that we have an octahedral lattice, namely $G = O_h$, where $O_h$ is the full octahedral symmetry group of order $48$, or group m3m in Hermann-Mauguin notation\cite{nespolo2017international}. However, much of the analysis remains unchanged for a different lattice system with a different point group, e.g. an hexagonal-closed pack (HCP) lattice with point group $D_{6h}$.

Then, from $N$ and the point group $O_h$, each family of lattice vectors can be defined by the action of the group $O_h$ on the lattice vector $\mathbf{n}$, i.e. $O_h\cdot \mathbf{n}$, where $\mathbf{n}\in N$. For example, the $\langle 110\rangle$ family is equal to $O_h\cdot [110] = \{\mathbf{T}[110]\;|\; \mathbf{T}\in O_h\} = \{ [110], [\overline{1}10], [101], \cdots \}$. Then, the set of unique lattice vector families is the set of all orbits of $N$ under $O_h$, i.e. $N/O_h$. For brevity, we will denote $\langle\mathbf{n}\rangle = O_h\cdot\mathbf{n}$.

Additionally, let $A$ be the set of alloying types in the system. Then, the unique unordered pairs of alloying types is $A^2/S_2 = \{\{\alpha, \alpha'\}\;|\;\alpha,\alpha'\in A\}$, where the inner braces denote a multiset, the outer braces denote a set, and $S_2$ is the permutation group on $2$ elements. Here, we quotient by $S_2$ to denote that the pairs are strictly unordered, i.e. $\{\alpha,\alpha'\} = \{\alpha',\alpha\}$.

Then, each unique dumbbell type is identifiable with an element of $(N/O_h)\times (A^2/S_2)$, i.e. a pair containing a lattice vector family and a pair of alloying elements. For example, for a BCC solid with the choice of $N$ above with elements Fe-Cr, the SIA types are $(\langle 110\rangle, \{\text{Fe}, \text{Cr}\})$, $(\langle 110\rangle, \{\text{Fe}, \text{Fe}\})$, $(\langle 110\rangle, \{\text{Cr}, \text{Cr}\})$, $(\langle 111\rangle, \{\text{Fe}, \text{Cr}\})$, $(\langle 111\rangle, \{\text{Fe}, \text{Fe}\})$, and $(\langle 111\rangle, \{\text{Cr}, \text{Cr}\})$. We can then compute the concentration of a given unique dumbbell type:

\begin{equation}\label{eq:self-interstitial-concentration}
    x_{\langle\mathbf{n}\rangle, \{\alpha, \alpha'\}}(\beta) = \sum_{\{\alpha'', \alpha'''\} = \{\alpha,\alpha'\}}\sum_{\mathbf{n}'\in \langle \mathbf{n}\rangle} x_{\mathbf{n}', \alpha'', \alpha'''}(\beta)
\end{equation}

i.e., we sum over defect types within the appropriate equivalence class. Importantly, the formulas above only apply for SIAs that are metastable or stable. Otherwise, within the context of coarse-graining continuous states into discrete ones as done above, a microstate for that particular SIA type is not well defined. This is important largely for two reasons:

\begin{itemize}
    \item The formulas above are wholly unapplicable to a SIA family that is mechanically unstable
    \item The formulas above must be treated with nuance in the case where SIAs are metastable or stable in some chemical environments and unstable in others
\end{itemize}

The former point is generally straightforward to address. Most simply, one can include all microstates that are possibly stable, and simply ignore these microstates if they relax significantly far away from their initial state. Even if some stable, but high-energy families are missed, e.g. $\langle 100\rangle$ SIA in BCC transition metals \cite{nguyen2006self}, these microstates will not significantly change any numerical results, since their Boltzmann factors will be neligible.

In contrast, the latter can be subtle. Changes in the local chemical environment can perturb the potential energy surface and partially break the symmetries associated with the orbits in $N/O_h$. For instance, in this study, a significant fraction of $\langle 111\rangle$ crowdions in $\text{Fe}_{1-x}\text{Cr}_x$ (with our choice of interatomic potential) misalign by more than $15^\circ$ upon relaxation, with misalignment clearly correlated with Cr concentration. However, this effect is a consequence of symmetry-breaking of configuration space, rather than a direct consequence of the proposed model. As such, this effect is strongly system-dependent, and will be discussed in detail later.

From this picture, we can compute the effective formation energy for a vacancy $E_\text{form}^{(v)}(\beta)$ as:

\begin{equation}\label{eq:vacancy-formation}
    E_\text{form}^{(v)}(\beta) = -\frac{\partial \ln x_v(\beta)}{\partial\beta}
\end{equation}

while, for a SIA, the effective formation energy $E_\text{form}^{(\alpha, \alpha', \langle\mathbf{n}\rangle)}(\beta)$, is:

\begin{equation}\label{eq:self-interstitial-formation}
    E_\text{form}^{(\langle\mathbf{n}\rangle, \{\alpha, \alpha'\})}(\beta) = -\frac{\partial \ln x_{\langle\mathbf{n}\rangle, \{\alpha, \alpha'\}}(\beta)}{\partial\beta}
\end{equation}

Here, the effective formation energies are interpretable as formation free energies for their corresponding defect subsystem. We can motivate this interpretation more abstractly by considering a discrete system consisting of a set of microstates $\mathcal{M}$, and a two-subsystem partition $\mathcal{M} = \chi^\circ \cup\chi$, where $\chi^\circ$ denotes a set of ``reference" microstates and $\chi$ denotes a set of defective microstates. For example, in this work, $\chi^\circ$ is the set of alloying elements and $\chi$ is either the singleton set consisting of a vacancy, or the set consisting of all equivalent microstates corresponding to some SIA type.

Without loss of generality, we can assume that the system is coupled to an energy and particle-exchanging reservoir. Then, the probability of a given microstate $\omega'\in\mathcal{M}$ is:

\begin{equation}
    p(\omega') = \frac{e^{-\beta \Tilde{E}_{\omega'}}}{\sum_{\omega\in\mathcal{M}} e^{-\beta\Tilde{E}_\omega}}
\end{equation}

where $\Tilde{E}_\omega$ denotes the Legendre-transformed energy, namely $\Tilde{E}_\omega = E_\omega - \mu N_\omega$, of a given microstate. Then, in the dilute limit, the probability of any one of the defect microstates $\omega\in \chi$ is:

\begin{equation}
    \begin{aligned}
        p(\chi) &= \sum_{\omega'\in\chi} p(\omega') = \frac{\sum_{\omega'\in \chi} e^{-\beta \Tilde{E}_{\omega'}}}{\sum_{\omega \in \mathcal{M}} e^{-\beta \Tilde{E}_\omega}}\\
        &= \frac{\sum_{\omega'\in \chi} e^{-\beta \Tilde{E}_{\omega'}}}{\sum_{\omega \in \chi} e^{-\beta \Tilde{E}_\omega} + \sum_{\omega\in \chi^\circ} e^{-\beta\Tilde{E}_\omega}}
    \end{aligned}
\end{equation}

Then, immediately, we can identify each sum as a Landau free energy of the subsystems $\chi$ and $\chi^\circ$, respectively labeled $\Omega_\chi$ and $\Omega_{\chi^\circ}$:

\begin{equation}
    \begin{aligned}
        p(\chi) = \frac{e^{-\beta \Omega_\chi}}{e^{-\beta \Omega_\chi} + e^{-\beta \Omega_{\chi^\circ}}} \approx e^{-\beta(\Omega_\chi - \Omega_{\chi^\circ})}
    \end{aligned}
\end{equation}

where we have made the approximation that the reference microstates are significantly more energetically stable than the defect microstates, i.e. $\Omega_{\chi^\circ} \ll \Omega_\chi$.

In our work, we compute conditional probabilities that depend on local chemical environments, namely $p(\chi \; | \; \eta)$, where $\eta$ denotes a local chemical environment. In this picture, the subsystem Landau free energies are functions of $\eta$:

\begin{equation}
    p(\chi\;|\;\eta) \approx e^{-\beta \left(\Omega_\chi(\eta) - \Omega_{\chi^\circ}(\eta)\right)}
\end{equation}

Then, $p(\chi)$ is simply a marginal distribution of $p(\chi\;|\;\eta)$, namely $p(\chi) = \langle p(\chi\;|\;\eta)\rangle_\eta$. So, the effective formation energy for the defect microstates is, as defined earlier:

\begin{equation}
    \begin{aligned}
        E_\text{form}^{(\chi)}(\beta) &= -\frac{\partial \ln \langle p(\chi\;|\;\eta)\rangle_\eta}{\partial\beta} \\
        &\approx -\frac{\partial}{\partial\beta}\ln\left\langle e^{-\beta\left(\Omega_\chi - \Omega_{\chi^\circ}\right)}\right\rangle\\
        &=\frac{\left\langle \Delta\Omega \; e^{-\beta\Delta\Omega}\right\rangle}{\langle e^{-\beta\Delta\Omega}\rangle}
    \end{aligned}
\end{equation}

where the $\eta$'s have been dropped in the final expression for brevity, the average implicitly means an average over local chemical environments, and $\Delta\Omega = \Omega_\chi - \Omega_{\chi^\circ}$.

The above expression shows a clear interpretation of the effective formation energy as defined in equations \eqref{eq:vacancy-formation} and \eqref{eq:self-interstitial-formation}. Namely, if $\Delta\Omega$ is the formation free energy for a particular defect type, which is taken to be a random variable over chemical environments, then the effective formation energy is the Boltzmann-weighted average of $\Delta\Omega$. In the case of a homogeneous solid, this quantity is simply the formation energy (neglecting vibrational states), since both subsystems of interests are non-degenerate, e.g. in the case of point defects, any dumbbell family is a singleton set, and the background reference state is a singleton set as well. In the case of a heterogeneous solid, this formation free energy includes the entropy difference between the two subsystems: namely the configurational entropy of the point defect, which is nonzero for SIA dumbbells, and the configurational entropy of the perfect crystal.

Note that the above derivation is additionally valid in a continuum picture, meaning one can include particular continuum quantities of interest for defects - most notably vibrational states - in the free energies $\Omega_\chi$ and $\Omega_{\chi^\circ}$, which are predicted to significantly affect properties of vacancies in Cu and Al at sufficiently high temperatures \cite{glensk2014breakdown}. However, this is computationally expensive, especially in the case of a combinatorally large number of defect types (as is the case for SIAs), so we only include Ising-like microstates in this work. Additionally, the above picture is valid for a system with more than 1 defect type, provided that the minimum-free-energy subsystem is significantly more stable than the others, i.e. $\Omega_{\chi^\circ} \ll \Omega_{\chi}$ for all $\chi\neq\chi^\circ$.

As in our prior work \cite{jeffries2025prediction}, this model is similarly amenable to other thermodynamic forces. Of particular interest, for example, is the formation volume of SIAs and vacancies in FeCr, which can be computed by introducing an external pressure $p$ - which induces a mechanical term in the microstate probabilities - and similarly differentating the log concentration of the defect type with respect to $\beta p$, rather than just $\beta$ as in equations \eqref{eq:vacancy-formation} and \eqref{eq:self-interstitial-formation}. Such a quantity is of interest, for example, in the study of irradiation swelling in ferritic steels, in which Frenkel pairs are reported to cause net swelling \cite{hu2022first}.

Lastly, we compute the chemical potentials using a Widom-like method, where we compute them from occupation energy differences \cite{10.1063/1.1734110}. We focus on binary alloys with alloying types $\alpha$ and $\alpha'$, yielding the two equations at $\SI{0}{K}$:

\begin{equation}
    \begin{aligned}
        \mu_\alpha - \mu_{\alpha'} &= \left\langle E_\sigma^{(\alpha)} - E_\sigma^{(\alpha')}\right\rangle \\
        x_\alpha\mu_\alpha + x_{\alpha'}\mu_{\alpha'} &= h
    \end{aligned}
\end{equation}

where $h$ is the enthalpy per atom of the reference configuration. Or, equivalently:

\begin{equation}\label{eq:chemical-potentials}
    \begin{aligned}
        \mu_\alpha &= h + x_{\alpha'} \left\langle E_\sigma^{(\alpha)} - E_\sigma^{(\alpha')}\right\rangle \\
        \mu_{\alpha'} &= h - x_{\alpha} \left\langle E_\sigma^{(\alpha)} - E_\sigma^{(\alpha')}\right\rangle
    \end{aligned}
\end{equation}

These chemical potentials are notably computed at $\SI{0}{K}$. The formulas at finite temperature are notably similar and trivial to include, namely swapping energy differences for free energy differences. However, for the sake of simplicity, i.e. since we already have the occupation energies $E_\sigma^{(\alpha)}$, we will use the $\SI{0}{K}$ chemical potentials in this work.
\section{Methods and Results}

Below, we compute vacancy and SIA concentrations and formation energies for two model systems, namely $\text{Fe}_{1-x}\text{Cr}_x$ and $\text{Cu}_{1-x}\text{Ni}_x$, for compositions $\SI{1}{at. \%} \leq x\leq \SI{20}{at. \%}$ with $\SI{1}{at. \%}$ spacing. The former was chosen due to the broad interest of ferritic steels in irradiated environments \cite{lechtenberg1985irradiation, maziasz1989formation, little1979void, tavassoli2014current, lu2008irradiation}, and the latter was chosen for two reasons, namely as an example of our methodology applied to FCC metals (as opposed to BCC ferritic steels), and due to the simplicity of its phase diagram, i.e. that $\text{Cu}_{1-x}\text{Ni}_x$ remains a solid solution for a large range of compositions and temperatures \cite{turchanin2007phase}.

For $\text{Fe}_{1-x}\text{Cr}_x$, we enumerate two families of SIA configurations, namely $\langle 110\rangle$ and $\langle 111\rangle$ SIAs, which are both known to be metastable \cite{olsson2007ab, fu2004stability, terentyev2008self}. For $\text{Cu}_{1-x}\text{Ni}_x$, we enumerate $\langle 100\rangle$ and $\langle 110\rangle$ SIAs, which are known to often be metastable in FCC metals \cite{ma2021nonuniversal}.

Occupation energies for each microstate were computed using the Large-scale Atomic/Molecular Massively Parallel Simulator (LAMMPS) \cite{thompson2022lammps} integration within the Atomic Simulation Environment (ASE) \cite{larsen2017atomic} package. In both cases, we use embedded-atom method (EAM) \cite{daw1984embedded} potentials that were developed to study point defect properties, respectively by Bonny et al. \cite{bonny2011iron} and Onat et al. \cite{onat2013optimized}.

First, we initialize a random solid solution and relax at a constant pressure of $\SI{0}{bar}$ \cite{tadmor1999mixed}. Note that we do not perform Monte Carlo swaps during this relaxation, meaning that our results are exclusively representative of the random phases of both systems studied in this work. We justify this choice for two main reasons: namely that we want to focus on our methodology and its interaction with random solid solution noise, and that the use of random solutions is a common choice for parameterizing phase field models \cite{piochaud2016atomic,badillo2015phase, rezwan2022effect, soisson2006kinetic, vizoso2021determination}. This is not an issue for $\text{Cu}_{1-x}\text{Ni}_x$, which remains a solid solution over the whole composition range for a large temperature range. However, $\text{Fe}_{1-x}\text{Cr}_x$ has a much more complex phase diagram, with significantly less solubility than $\text{Cu}_{1-x}\text{Ni}_x$. Meaning that, for example, for high $x$ and sufficiently low temperatures, our results for $\text{Fe}_{1-x}\text{Cr}_x$ are importantly not that of an equilibrium phase, and must be considered with care.

Using this reference configuration, we randomly sample lattice sites $\sigma$ without replacement. To then compute the occupation energies of the alloying types $E_\sigma^{(\alpha)}$, the vacancy $E_\sigma^{(v)}$, and the SIAs $E_\sigma^{(\mathbf{n},\alpha,\alpha')}$, we place each alloying or defect type at the lattice site, and minimize the energy at constant pressure. 

From the atomic occupation energies $E_\sigma^{(\alpha)}$, we compute chemical potentials as a function of composition following equation \eqref{eq:chemical-potentials}. We have included a plot of the chemical potentials as a function of composition in the Supplementary Materials.

These energies and chemical potentials are then fed into equations \eqref{eq:vacancy-concentration} and \eqref{eq:self-interstitial-concentration} for $\SI{7.7}{eV^{-1}}\leq\beta\leq \SI{40.0}{eV^{-1}}$, approximately corresponding to temperatures $\SI{300}{}$-$\SI{1500}{K}$. Then, following equations \eqref{eq:vacancy-formation} and \eqref{eq:self-interstitial-formation}, we compute the respective formation energies for the defect of type $d$ via a central finite difference:

\begin{equation}
    E_\text{form}^{(d)}(\beta_n) \approx -\frac{\ln x_d(\beta_{n+1}) - \ln x_d(\beta_{n-1})}{\beta_{n+1} - \beta_{n-1}}
\end{equation}

We then smooth the formation energy across composition $x$ using the Savitzky-Golay filter \cite{schafer2011savitzky} with a window length of $9$ and a polynomial order of $3$.

We note that, during relaxation, SIAs in a solid solution need not remain along their ideal crystallographic axis. In extreme cases, a SIA will relax far away from its original crystallographic axis, effectively yielding mislabeling of SIA microstates. In this work, we label a SIA microstate as misaligned if the initial and final axes along the SIA differ by more than $15^\circ$, and ignore the misaligned SIA microstates.

Below, we present our calculated effective formation energies for each system of interest over the aforementioned composition and temperature ranges. In each effective formation energy plot, we show $E_\text{form}$ as a function of composition $x$ for multiple temperatures in the form of a heatmap, where each contour on the heatmap represents an isotherm. For completeness, we have included similar heatmaps for the defect concentrations, of which the formation energies are derived from, within the Supplementary Materials.

\subsection{FeCr}

\begin{figure}[H]
    \centering
    \includegraphics[width=\linewidth]{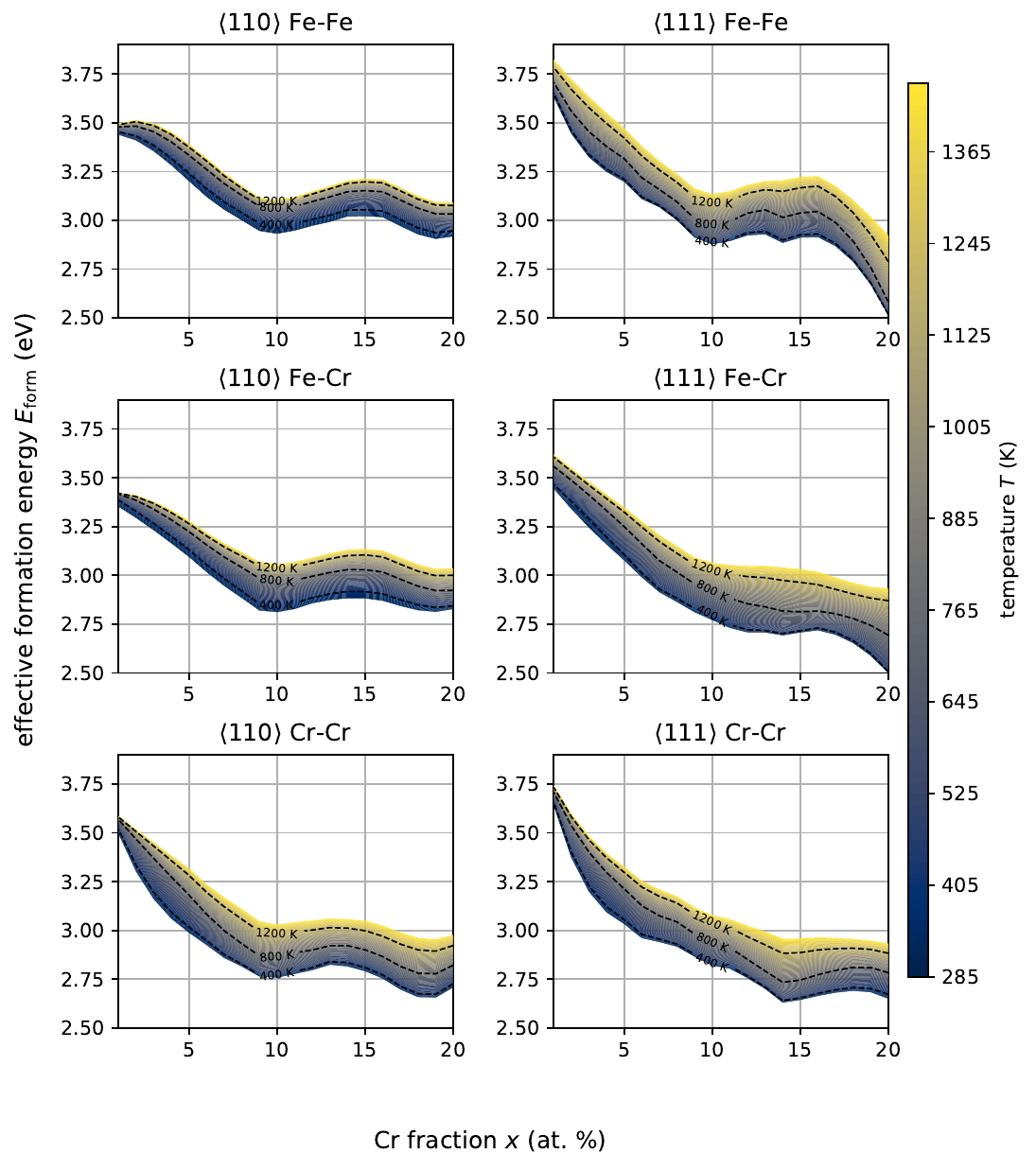}
    \caption{Effective formation energies for each SIA dumbbell type in $\text{Fe}_{1-x}\text{Cr}_x$ as a function of composition and temperature.}
    \label{fig:fe-cr-SIAs}
\end{figure}

\begin{figure}[H]
    \centering
    \includegraphics[width=\linewidth]{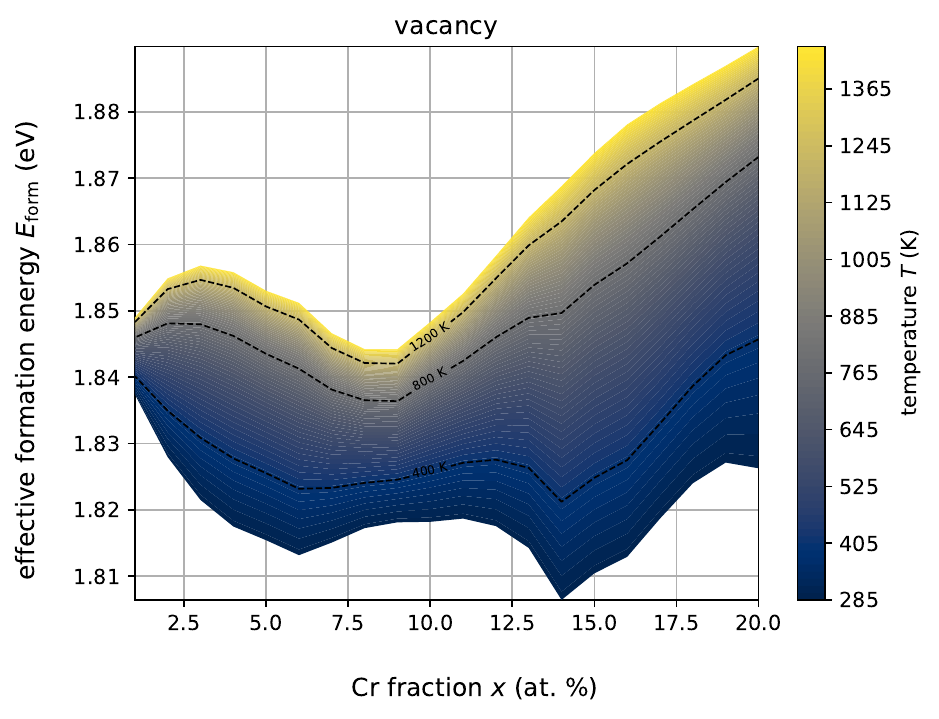}
    \caption{Effective formation energies for vacancies in $\text{Fe}_{1-x}\text{Cr}_x$ as a function of composition and temperature.}
    \label{fig:fe-cr-vacancies}
\end{figure}

First, we present results on the effective formation energies for the $\text{Fe}_{1-x}\text{Cr}_x$ system using the aforementioned potential developed by Bonny et al. \cite{bonny2011iron} for the various defect types, namely the $\langle 110\rangle$ and $\langle 111\rangle$ SIAs in Figure~\ref{fig:fe-cr-SIAs} and vacancies in Figure~\ref{fig:fe-cr-vacancies}.

For the SIAs, we find that, in pure Fe, the different SIA types have significantly different stabilities. Namely, the $\langle 110\rangle$ dumbbells are all lower than the $\langle 111\rangle$ crowdions in effective formation energy, and the Fe-Cr SIAs in dilute alloys are lower in formation energy than both of their corresponding Fe-Fe and Cr-Cr dumbbells. This is rather unsurprising, and consistent with prior literature on SIA dumbbells in dilute Fe-Cr \cite{senninger2016modeling, olsson2007ab, terentyev2008migration}.

However, interestingly, this ordering does not necessarily hold over the entire composition range. Namely, as the solid solution becomes more concentrated in Cr, we see that formation energies vary more with temperature, which we attribute to the configurational entropy of each SIA subsystem. This has effects which break the ordering in dilute FeCr alloys. Notably, we predict that the formation energy of a $\langle 111\rangle$ Fe-Cr dumbbell is slightly lower than that of a $\langle 110\rangle$ Fe-Cr dumbbell along the $\SI{800}{K}$ contour at $10\%$ Cr. Additionally, we predict that the most stable SIA types at low temperatures (near $\SI{400}{K}$) are $\langle 111\rangle$ Fe-Fe and Fe-Cr, in stark contrast with the most stable types for dilute alloys described above.

We additionally see that the effective formation energy for a vacancy varies more with temperature for more concentrated solutions, however the effect is significantly less drastic, with the effective formation energy varying by less than $\SI{0.1}{eV}$ over the entire composition and temperature range. This lesser variation is largely consistent with our picture, namely that an SIAs has more configurational entropy than a vacancy, which has no configurational entropy. We note that, although the trend for vacancies looks more complex than the trend for SIAs over our chosen composition range, the heatmap for vacancies has a much smaller set of limits over the formation energy axis, and attribute the corresponding small fluctuations largely to noise in our statistics.

\subsection{CuNi}

\begin{figure}[H]
    \centering
    \includegraphics[width=\linewidth]{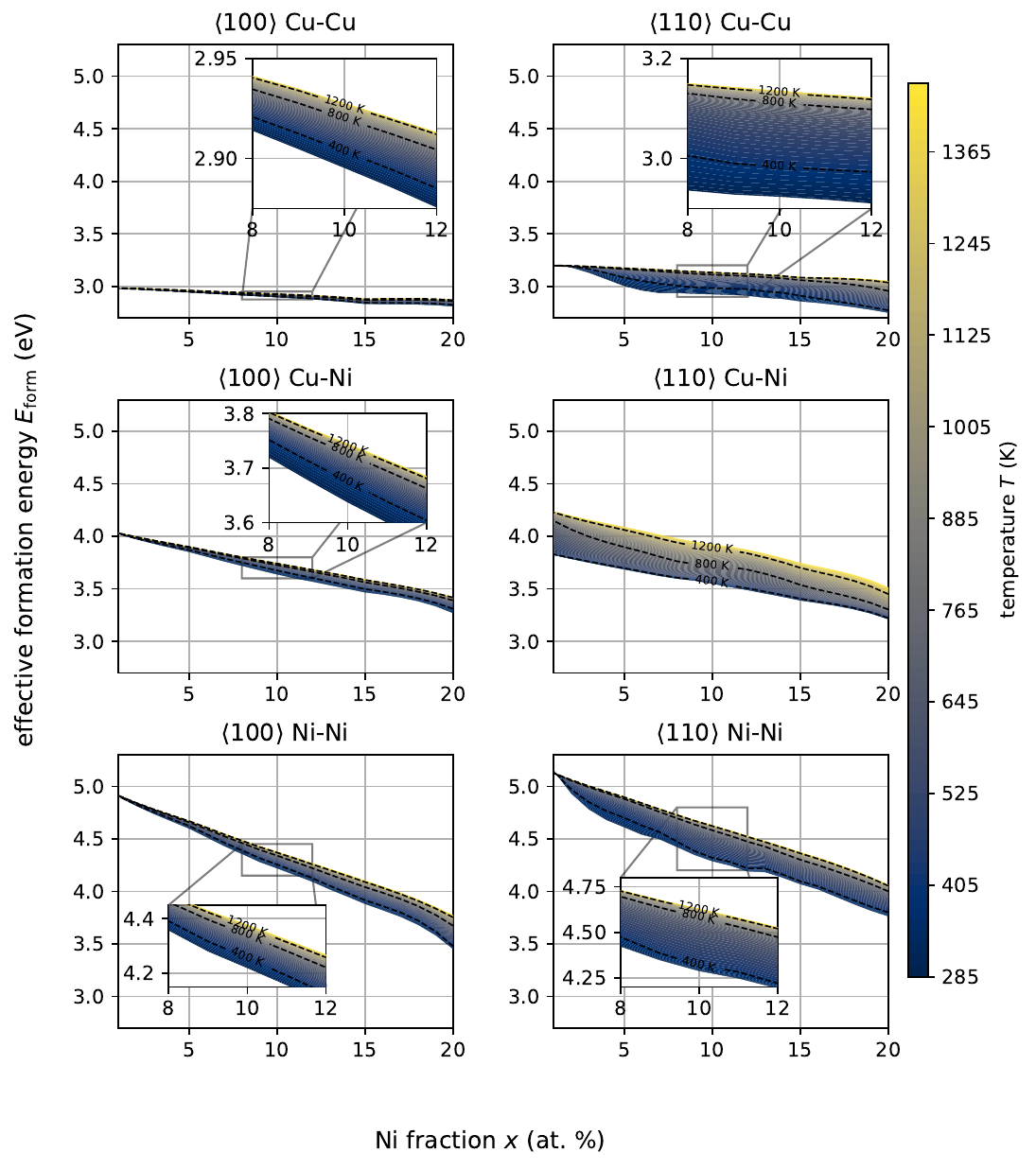}
    \caption{Effective formation energies for each SIA dumbbell type in $\text{Cu}_{1-x}\text{Ni}_x$ as a function of composition and temperature.}
    \label{fig:cu-ni-SIAs}
\end{figure}

\begin{figure}[H]
    \centering
    \includegraphics[width=\linewidth]{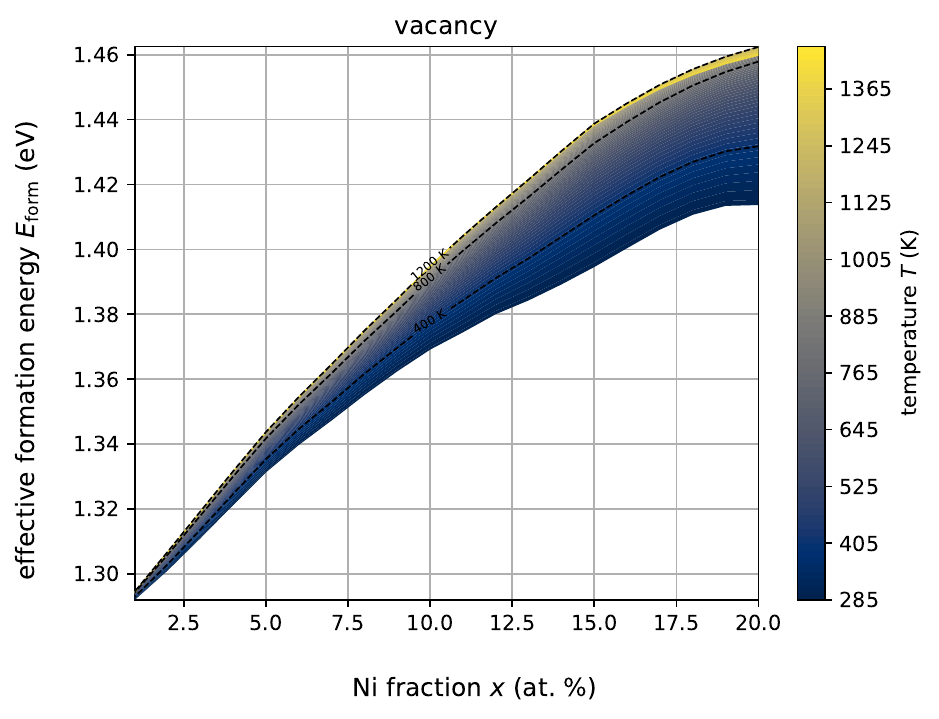}
    \caption{Effective formation energies for vacancies in $\text{Cu}_{1-x}\text{Ni}_x$ as a function of composition and temperature.}
    \label{fig:cu-ni-vacancies}
\end{figure}

Second, we present results on the effective formation energies for the $\text{Cu}_{1-x}\text{Ni}_x$ system using the aforementioned potential developed by Onat et al. \cite{onat2013optimized} for the various defect types, namely the $\langle 100\rangle$ and $\langle 110\rangle$ SIAs in Figure~\ref{fig:cu-ni-SIAs} and vacancies in Figure~\ref{fig:cu-ni-vacancies}.

For the SIAs, in contrast with the $\text{Fe}_{1-x}\text{Cr}_x$ case, we predict that the ordering of the stability of SIAs remains largely constant over the composition range, namely that the stability of the SIAs is largely a function of chemistry rather than orientation, i.e. that the formation energies are ordered by their chemical types, namely Cu-Cu $<$ Cu-Ni $< $ Ni-Ni. Interestingly, we also see that the variation of effective formation energy over temperature is significantly smaller for the $\langle 100\rangle$ dumbbells. This is consistent with our picture of SIA dumbbells as equivalence classes over orbits, since $|O_h\cdot[110]| = 2|O_h\cdot[100]|$, i.e. the $\langle 110\rangle$ family of lattice vectors is twice the size of the $\langle 100\rangle$ family, yielding relatively higher configurational entropy.

Similar to the $\text{Fe}_{1-x}\text{Cr}_x$ case, we see larger variations of the effective formation energy over temperature as the solid solution becomes more concentrated. However, in contrast to the $\text{Fe}_{1-x}\text{Cr}_x$ case, said variation is more systematic and larger in magnitude. Since the vacancy microstate equivalence class has zero entropy by definition in this work, we attribute this solely to the high mixing entropy of the underlying lattice, which is consistent with the high stability of the solid solution phase in Cu-Ni alloys \cite{turchanin2007phase}.

\subsection{Symmetry breaking}

\begin{figure}[H]
    \centering
    \includegraphics[width=\linewidth]{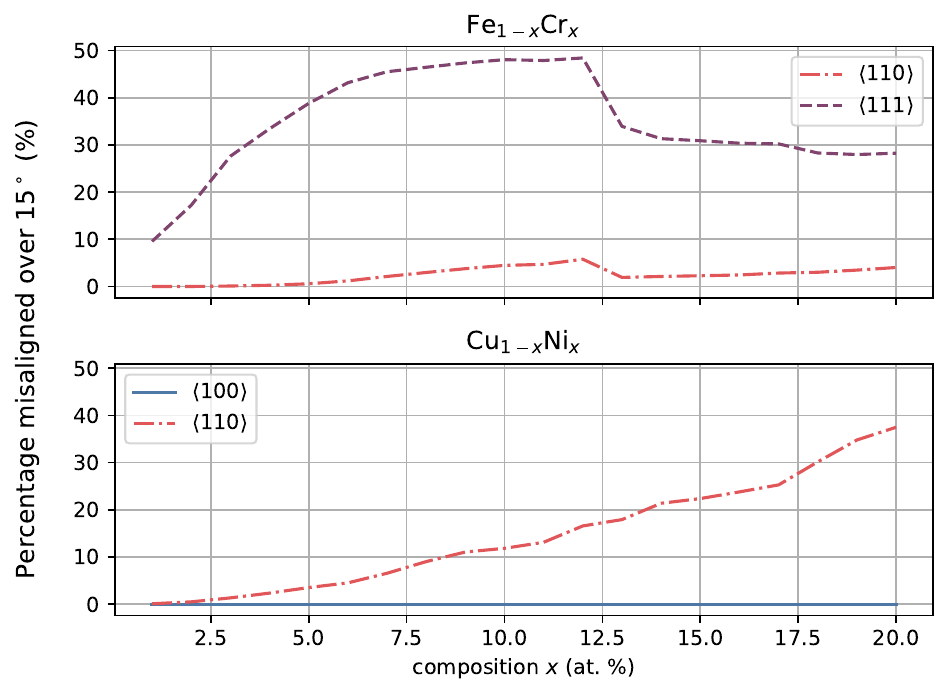}
    \caption{Misalignment of dumbbell microstates in both systems of interest. A microstate is labeled misaligned if the dumbbell's final axis differs from its initial axis by more than $15^\circ$ upon relaxation.}
    \label{fig:misalignment}
\end{figure}

In both systems studied in this work, we see an interesting effect, namely that some SIAs significantly misalign with respect to their ideal axis $\mathbf{n}$ upon relaxation. In Figure~\ref{fig:misalignment}, we showcase the proportion of microstates that become significantly misaligned, of which we have chosen an arbitrary threshold of $15^\circ$. For reference, this angle is close to $\angle([100], [410]) \approx 14^\circ$, $\angle([110], [210]) \approx 18^\circ$, and $\angle([111], [221]) \approx 16^\circ$, where $\angle(\mathbf{n}, \mathbf{m})$ is the angle between vectors or axes $\mathbf{n}$ and $\mathbf{m}$.

In $\text{Fe}_{1-x}\text{Cr}_x$, we see that $\text{Cr}$ concentration leads to increasingly misaligned SIAs up until about $12\%$ Cr, with nearly half of the $\langle 111\rangle$ microstates becoming misaligned. A smaller fraction of $\langle 110\rangle$ microstates become misaligned with increasing Cr concentration, which is consistent with Klaver et al. \cite{klaver2007interstitials} and Wr{\'o}bel et al. \cite{wrobel2021elastic}, who both predicted that, from density functional theory (DFT) calculations, $\langle 110\rangle$ Cr-Cr SIAs tend to relax to $\langle 221\rangle$ directions as the local concentration of Cr increases.

In contrast, we find that $\langle 110\rangle$ SIAs in $\text{Cu}_{1-x}\text{Ni}_x$ are monotonically misaligned by Ni concentration, with no $\langle 100\rangle$ SIAs significantly misaligning. This effect is similarly quite significant for the $\langle 110\rangle$ microstates, with nearly $40\%$ of SIAs misaligning at $20\%$ Ni.

In both cases, we see that variations in local environment effectively break the point-group symmetry as described in the formulation. Previously, we defined each defect type as an equivalence class induced via orbits of the lattice's point group $G$. More precisely, two SIAs with the same chemical types $\alpha$ and $\alpha'$ pointing along directions $\mathbf{m}$ and $\mathbf{n}$ are equivalent if and only if the two orbits of both vectors are the same, namely $\mathbf{m}\sim\mathbf{n}$ if and only if $G\cdot\mathbf{m} = G\cdot\mathbf{n}$.

\begin{figure}[H]
    \centering
    \includegraphics[width=\linewidth]{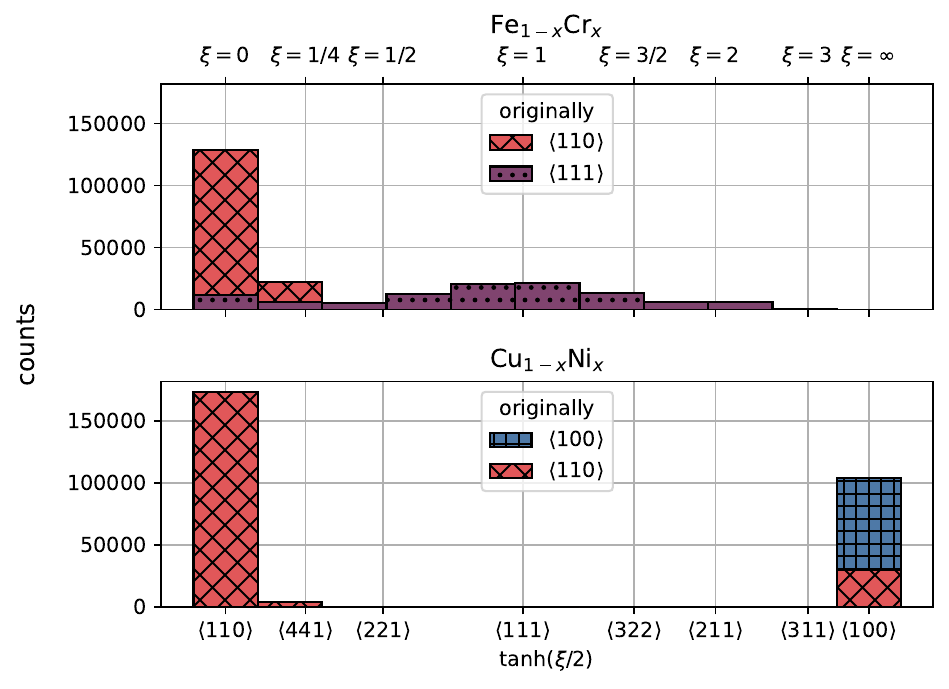}
    \caption{Distributions of $\tanh(\xi/2)$ for the dumbbell microstates, labeled by the original dumbbell direction.}
    \label{fig:misalignment2}
\end{figure}

However, as evident above, it need not be true that the full orbit is stable. We see that compositional fluctuations in local chemical environments significantly shift local minima in a given SIA dumbbell's configuration space, whose new low-symmetry minima need not be related by any residual point-group symmetry. Specifically, a large proportion of the $\langle 111\rangle$ Fe-Cr SIA microstates in this work relax to $\langle 110\rangle$-like states. As done in prior literature \cite{ma2019symmetry, wrobel2021elastic}, we quantify this by identifying each SIA with a family $\langle 11\xi\rangle$, where $\xi$ is a continuously varying parameter between $0$ and $\infty$. In Figure~\ref{fig:misalignment2}, we showcase distributions of $\tanh(\xi/2)$ over the whole composition range for both systems. We plot $\tanh(\xi/2)$ rather than $\xi$ to ensure that the range of values is finite, i.e. $\tanh(\xi/2) = 0$ for $\langle 11\xi\rangle = \langle 110\rangle$ and $\tanh(\xi/2) = 1$ for $\langle 11\xi\rangle = \langle 100\rangle$. Here, the factor of $1/2$ ensures that the SIA families of interest, namely the $\langle 110\rangle$, $\langle 111\rangle$, and $\langle 100\rangle$ families, are roughly equally spaced apart, since $\tanh(1/2)\approx 1/2$. This is entirely a visualization technique, and the choice of $\tanh$ is arbitrary, motivated by $\tanh$'s popular choice as an activation function. The choice of a misalignment angle threshold of $15^\circ$ is also arbitrary, and will likely be a function of the application of interest. For example, this choice of $15^\circ$ means that $\langle 221\rangle$ and $\langle 322\rangle$ dumbbells are counted as $\langle 111\rangle$ configurations, while $\langle 311\rangle$ dumbbells are counted as $\langle 100\rangle$ dumbbells. A full visualization of the effects of this choice is included in the Supplementary Materials.

In $\text{Fe}_{1-x}\text{Cr}_x$, we see that the $\langle 110\rangle$ SIA dumbbells continuously bleed away from the $\langle 110\rangle$ label, which is unsurprising, and consistent with the symmetry-breaking description above, namely that only a small proportion of them significantly misalign. However, the case for the $\langle 111\rangle$ crowdions is more drastic: a significant portion of microstates bleed into the nearby, lower symmetry $\langle 221\rangle$-like and $\langle 211\rangle$-like bins. 

We note that this spreading does not invalidate our results presented on formation free energy in either system, since severely misaligned microstates were discarded prior to performing any statistics, but rather makes the results more nuanced: the formation free energies calculated are not broadly assignable to the entire orbit, but rather the restricted number of stable microstates within the orbit. This is significant for understanding the physical orientations that SIAs can take in concentrated solutions: although there are composition and temperature regions where $\langle 110\rangle$ dumbbells have lower formation energy than $\langle 111\rangle$, the number of metastable $\langle 110\rangle$ basins on the defect potential energy surface can be significantly restricted as a function of local chemical environment. This is particularly important for studying, for example, SIA migration pathways, e.g. if a particular migration pathway relies on stability of any $\langle 111\rangle$ orientation that is significantly symmetry-broken, then that pathway is inaccessible. Furthermore, such a conclusion suggests that care must be taken in atomistic kinetic Monte Carlo models, since the number of $\langle 110\rangle$ basins seems to strongly correlate with Cr concentration.

Similarly, in $\text{Cu}_{1-x}\text{Ni}_x$, we see that a non-negligible amount of $\langle 110\rangle$ SIAs relax to $\langle 100\rangle$-like states, which has the same ramifications as above. However, this effect appears less drastic, i.e. only a minority of these states relax to $\langle 100\rangle$-like states. In contrast to $\text{Fe}_{1-x}\text{Cr}_x$, we see that nearly zero $\langle 100\rangle$ dumbbell microstates relax to a different bin, meaning that symmetry is not broken for $\langle 100\rangle$ dumbbells for almost any local chemical environment in this study.

Broadly, we see that chemical disorder has the potential to non-negligibly break the orientational symmetry of a SIA configuration space without inducing a reduced symmetry group, which must be carefully treated for in any atomistic simulation involving SIAs, e.g. kinetic Monte Carlo simulations of irradiated alloys, in which SIA migration mechanisms must be enumerated prior to SIA hopping \cite{senninger2016modeling}. Additionally, this effect is conceptually consistent with broader literature regarding symmetry breaking in random alloys, e.g. Jahn-Teller splitting in $\alpha$-Ti alloys \cite{hu2020unconventional, chen2023spontaneous}.
\section{Limitations}

A primary limitation of the methodology above is that, in sampling lattice sites independently of one another, we implicitly assumed to be within a very dilute regime where point defects are spatially uncorrelated with respect to other point defects. We expect that this is unproblematic for studying quantities appropriate for this single-site picture, e.g. formation free energy of a point defect. However, as in our prior work \cite{jeffries2025prediction}, this single-site picture is likely insufficient for studying effects of correlated point defects, e.g. SIA clusters and loops, which are of particular interest in irradiation applications, e.g. for studying irradiation creep induced by dislocations\cite{onimus2021irradiation} and swelling induced by defect clusters \cite{terentyev2005correlation, chen2019investigation}.

We note that this is not a fault of a statistical picture, but rather that we sample using single sites as approximations for local chemical environments. For higher-dimensional defects, we expect that it is instead sufficient to sample two sub-regions of configuration space: namely the sub-region of configurations containing said defect, and the sub-region of configurations that are defect-less. From these samples, one could compute free energies using thermodynamic integration techniques, including but not limited to adiabatic switching\cite{de1996einstein, freitas2016nonequilibrium}, finite difference thermodynamic integration \cite{mezei1987finite}, and Bennet's acceptance ratio method \cite{bennett1976efficient}.

An additional limitation is finite-size effects. Especially for point defects, finite-size artifacts are problematic for \textit{ab-initio} methods, which are particularly constrained by system size \cite{erhart2006first, maier2000point, kumagai2014electrostatics}. This effect is expected to be particularly strong for point defects like SIAs, which induce a long-range elastic field \cite{burr2017importance}. As with any \textit{ab-initio} calculation, we recommend that each result at a single system size is considered only within the context of a convergence study.

A third limitation of our results is the explicit exclusion of vibrational microstates. This assumption is likely insufficient for describing point defects in large class of materials at high temperatures. For example, quasi-harmonicity and anharmonicity are both known to play a crucial role in vacancy formation in both Al and Cu \cite{glensk2014breakdown}. Similarly to our proposed solution for correlated defects, we believe that such effects are amenable by sampling vibrational microstates. However, this is computationally expensive, and is therefore saved for a future work.

Lastly, our results depend on the quality of the chosen interatomic potentials. This is potentially problematic for $\text{Fe}_{1-x}\text{Cr}_x$, whose SIA dumbbells are known to be significantly affected by magnetic effects \cite{wrobel2021elastic}, of which a classical interatomic potential cannot fully capture. Klaver et al. found significant inconsistencies in interstitial formation energies in Fe-Cr when using a set of EAM-style interatomic potentials fit to defect energetics, with large errors present in Cr-rich regions \cite{klaver2016inconsistencies}. 

This choice is largely driven by the computational inexpensiveness of an interatomic potential over an \textit{ab-initio} method. Including microstates over the entire composition range, each set of formation energy curves per system includes $\sim 10^5$ microstates, which would entail significant computational expense had we chosen an \textit{ab-initio} method. We expect that sampling every possible microstate is unnecessary, and that one can instead cleverly sample the defect potential energy surface to reduce computational cost, e.g. with a Monte Carlo method \cite{lee2009formation, sugimoto2014hydrogen, alfe2005schottky}. This, however, is out of the scope of this study, and is similarly saved for a future work.

\section{Conclusions}

In this work, we extend our prior work on a statistical framework for point-like defects in concentrated solid solutions to self-interstitial atoms (SIA) by constructing equivalence classes induced by their respective symmetries. From this construction, we compute defect formation free energies that implicitly include the configurational entropy of each SIA type. We then apply this extension to compute SIAs formation free energies in two systems of interest: $\text{Fe}_{1-x}\text{Cr}_x$ and $\text{Cu}_{1-x}\text{Ni}_x$. From our results, we predict that the properties of SIAs can significantly differ for very modest solute concentration, owing to the solute-induced symmetry-breaking of the defect free energy surface. Notably, we predict that, in $\text{Fe}_{1-x}\text{Cr}_x$, the presence of Cr yields significant misalignment of $\langle 111\rangle$ crowdions, and that the $\langle 111\rangle$ SIAs that do persist are often more thermodynamically favorable than $\langle 110\rangle$ dumbbells over a significant range of compositions and temperatures. This prediction suggests that $\langle 111\rangle$ crowdions should not be excluded in the modeling of concentrated $\text{Fe}_{1-x}\text{Cr}_x$, and that care needs to be taken in atomistic models which might assume the existence of particular defect types at each lattice site, e.g. atomistic kinetic Monte Carlo models.
\section{Data Availability}

A brief tutorial for computing our calculated formation energies and concentrations for the Cu-10\%Ni case is available on GitHub \cite{github_repo}. The full simulation data used and computed within this study will be made available upon request.
\section{Acknowledgements}

J.J. and E.M. acknowledge support from the U.S. Department of Energy, Office of Basic Energy Sciences, Materials Science and Engineering Division under Award No. DE-SC0022980. H.L., A.E. and E.M. acknowledge support from the U.S. Department of Energy, Office of Science, Office of Fusion Energy Sciences under grant number DE-SC0024515.

Additionally, this material is based on work supported by the National Science Foundation under Grant Nos. MRI\# 2024205, MRI\# 1725573, and CRI\# 2010270 for allotment of compute time on the Clemson University Palmetto Cluster.
\section{Disclaimer}

Any opinions, findings, and conclusions or recommendations expressed in this material are those of the author(s) and do not necessarily reflect the views of the National Science Foundation.

\bibliography{bibfile.bib}

\end{document}